\documentclass{iaus}
\usepackage{graphicx}

\title[Modelling stellar populations at high redshift] 
{Modelling stellar populations at high redshift}

\author[Claudia Maraston]   
{Claudia Maraston$^1$
}

\affiliation{$^1$ICG-University of Portsmouth, \\ 
PO13FX, Portsmouth, United Kingdom\\ email: {\tt claudia.maraston@port.ac.uk} \\[\affilskip]}

\pubyear{2011}
\volume{xxx}  
\pagerange{00--00}
\jname{Title of your IAU Symposium}
\editors{A.C. Editor, B.D. Editor \& C.E. Editor, eds.}
\begin{document}

\maketitle
\begin{abstract}
Stellar populations carry information about the formation of galaxies and their evolution up to the present epoch. A wealth of observational data are available nowadays, which are analysed with stellar population models in order to obtain key properties such as ages, star formation histories, stellar masses. Differences in the models and/or in the assumptions regarding the star formation history affect the derived properties as much as differences in the data. I shall review the interpretation of high-redshift galaxy data from a model perspective. While data quality dominates galaxy analysis at the highest possible redshifts ($z>5$), population modelling effects play the major part at lower redshifts. In particular, I discuss the cases of both star-forming galaxies at the peak of the cosmic star formation history as well as passive galaxies at redshift below 1 that are often used as cosmological probes. Remarks on the bridge between low and high-$z$ massive galaxies conclude the contribution. 
\end{abstract}
\firstsection 
\section{Stellar evolution as a function of redshift}
The stellar content of galaxies evolves because of stellar evolution and because of galaxy evolution. While the second process is the one we would like to understand, the first one is in principle relatively well-known through the robust theory of stellar evolution. By exploiting this theory we can calculate the theoretical spectral energy distribution of populations of stars with arbitrary ages, chemical compositions, initial mass functions, so-called stellar population models, and compare them to observational data. 
The analysis of galaxy stellar populations is a crucial constrain to galaxy formation theories and cosmological models, because stars evolve on timescales that are ruled by nuclear physics hence are independent of the cosmological timescale. However, the physics of the adopted models and the assumptions made affect the derived properties as much as the use of different data sets.  

Stellar population models are calculated assuming (1) stellar 
evolution models which provide the energetics at given stellar mass; 
(2) stellar spectra for distributing the energetics to the various 
wavelengths; and (3) a numerical algorithm to calculate the 
integrated spectral energy distribution (SED). Over the last two decades these models have matured substantially. They now include all major stellar evolutionary phases and have a high spectral resolution (see Maraston 2011a for a recent review). 
\begin{figure}
\centerline{
\scalebox{0.3}{
\includegraphics{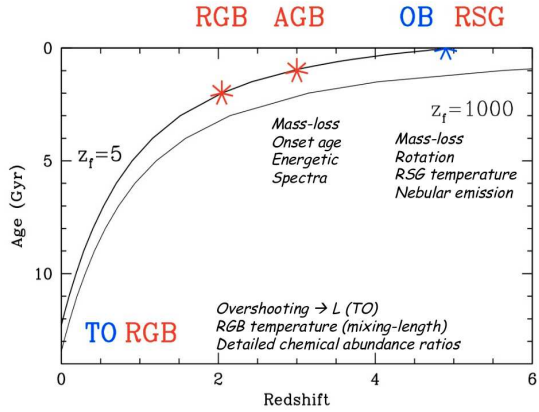}}}
\caption{The timeline of evolution of a passive population formed at $z\sim5$, marking the epochs at which major evolutionary phases dominate the light, and the uncertainties in the modelling.}
\label{starevol}
\end{figure}
Figure~\ref{starevol} shows the timeline of evolution of a population forming at redshift $z_{\rm f}\sim5$, in the idealised case of passive evolution, with highlighted the epochs at which major stellar evolutionary phases dominate the light. The uncertainties affecting the modelling of each phase are labelled. We use this Figure to discuss modelling of galaxy populations at various redshifts.
\subsection{The highest redshifts}
The youngest galaxies at the earliest epochs are dominated by O,B-type stars and Red Supergiants (RSG), and maybe strongly affected by dust reddening. Uncertainties in stellar evolution include the size of convective core-overshooting, the effect of stellar rotation, mass-loss, which determines the temperature of the RSG phase (e.g. Chiosi \& Maeder~1986). The consideration of nebular emission is also important at very young ages (e.g. Leitherer et al.~1999).  
\begin{figure}
\centerline{
\scalebox{0.45}{
\includegraphics{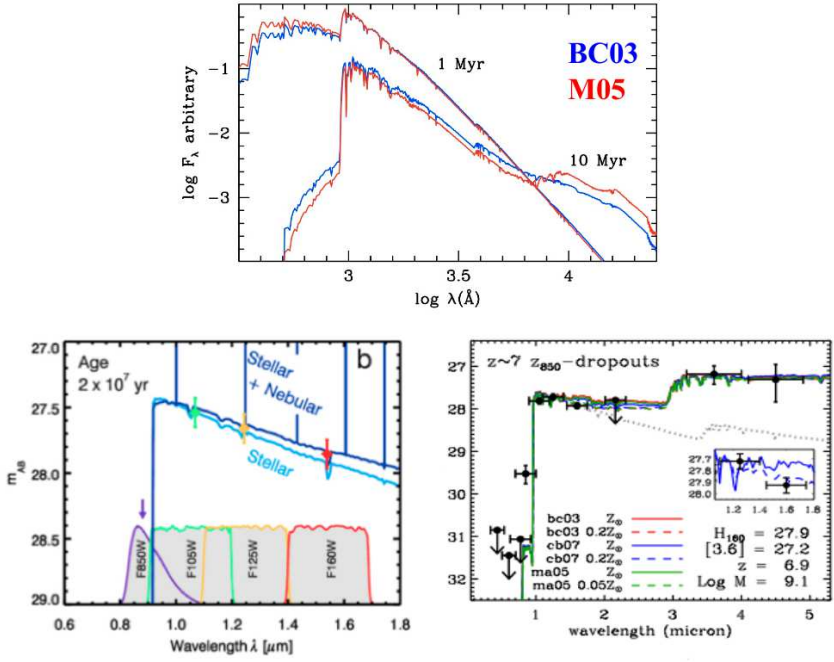}}}
\caption{Stellar population modelling of very high-redshift galaxies. The upper panel shows very young SSP models by Maraston (2005, red) and Bruzual \& Charlot (2003, blue), which are based on different stellar evolutionary tracks. The lower, left-hand panel shows the effect of nebular emission (visualised using a figure from Robertson et al. 2010). The lower, right-hand panel shows model fits to redshift 7 galaxy data, in which different population models are used (from Labb\`e et al. 2010).}
\label{red7}
\end{figure}
Figure~\ref{red7} collects plots related to the stellar population modelling of the highest redshift galaxies. The upper panel shows the difference between young models - 1 and 10 Myr - according to different codes specifically adopting different stellar tracks (see caption). While the 1 Myr models agrees very well, when the RSG phase kicks in ($\sim~10$~Myr) the Maraston (2005) models (red lines) have a higher flux in the rest-frame near-IR, due to the reddest RGS phase in the Geneva tracks (see also Marigo et al. 2008). This difference maybe of relevance when also nebular emission is included in the modelling, as both enhance the near-IR flux (as visualised in the lower left panel). 

All in all - however - in the highest redshift regime when galaxy ages are smaller than $\sim~100$ Myr, the physical interpretation of galaxy data is dominated by the data quality, rather than by the adopted population model, as models are quite similar. As an example, the lower right panel shows model fits to redshift 7 galaxy data, in which both Maraston-type and Bruzual \& Charlot-type of models are used. The fits and derived galaxy parameters are very similar (see also Sm\^olc\`ic et al.~2011). Besides the data quality, the adopted star formation history may matter, as we shall discuss below. 

\subsection{Redshift $3 - 1$: star formation and population model effects}
Climbing down the redshift ladder, as galaxies get older and the allowed range in ages is larger, the adopted stellar population prescriptions matter more, as well known and discussed in the literature by a large numbers of articles (see e.g. Maraston et al. 2011 for a summary). This is mostly due to the different prescriptions for the Thermally-Pulsating Asymptotic Giant Branch phase (TP-AGB), the inclusion of which enhances the near-IR flux of a stellar population model at ages around 1 Gyr, which are typically found in redshift 2-3 galaxies. In particular, due to this effect younger ages are obtained with the Maraston models, hence lower stellar masses, because the model SED becomes red at a younger age. This result is usually obtained when the SED fit includes at least some data in the rest-frame near-IR, which is where the red TP-AGB stars emit most of their light (see e.g. Maraston et al. 2006 for a detailed discussion). 

\begin{figure}
\centerline{
\scalebox{0.3}{
\includegraphics{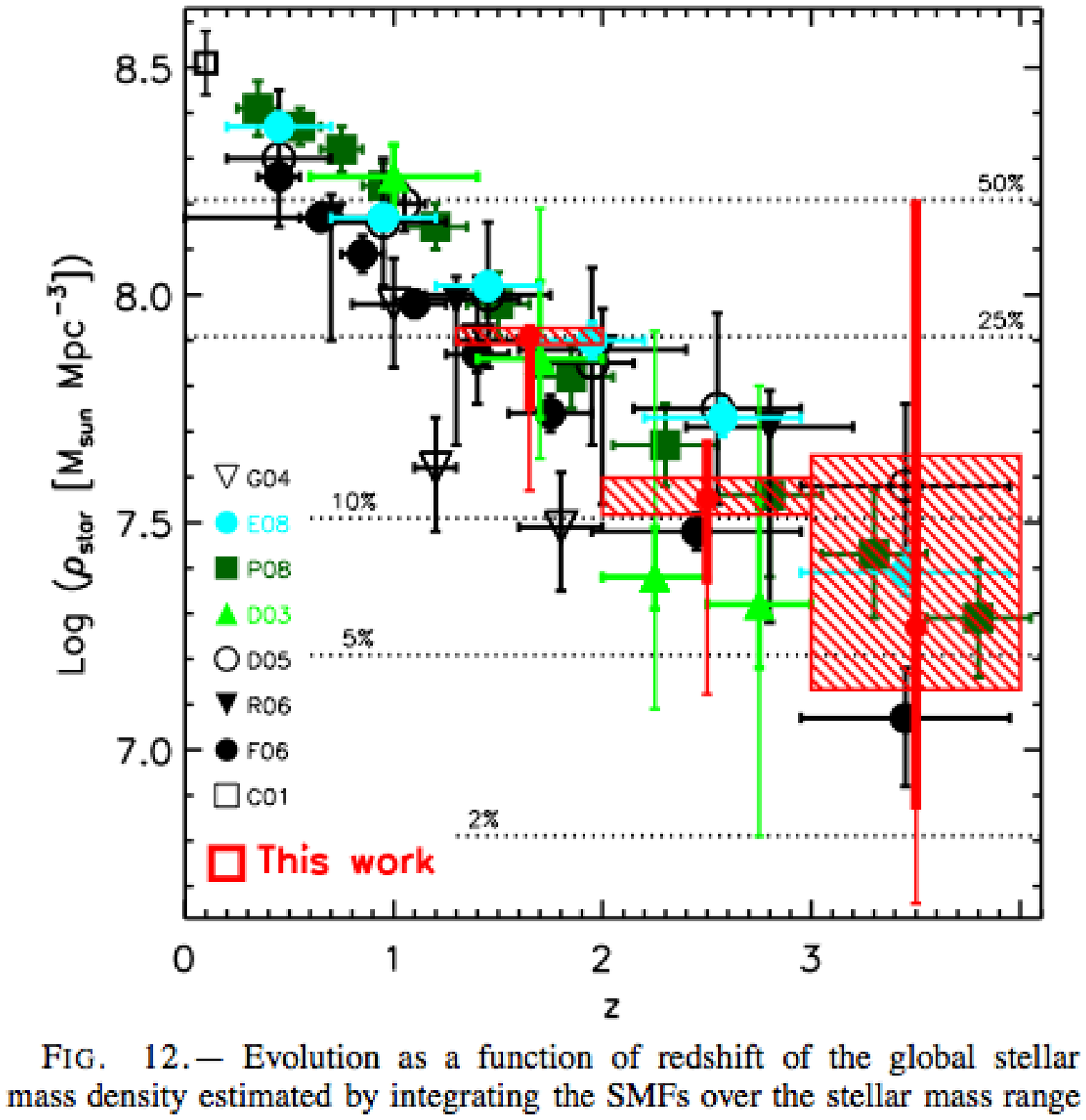}}}
\caption{The cosmic stellar mass density as a function of redshift. Stellar masses are derived by applying stellar population models to galaxy spectro-photometric data. Red squares represent galaxy derived masses where the uncertainties due to different stellar population models, stellar initial mass functions, star formation histories were taken into account. From Marchesini et al. 2009.}
\label{mf}
\end{figure}

What we stress here is that different stellar population models shape the empirical tracing of galaxy evolution. As an illustrative example, Figure~3 shows the galaxy stellar mass function as a function of redshift. Galaxy stellar masses are obtained fitting stellar population models to galaxy data, hence the choice of models and parameters drive the results. This is what the red squares represent, namely the variation in the derived stellar mass of a galaxy due to various population models (e.g. including or not a sizable TP-AGB phase), IMFs and star formation histories. It seems quite hard to set meaningful constrains to galaxy formation models if one takes these large variations at face value (see also reviews by M. Dickinson and O. Le F\`evre, this volume). 

\begin{figure}
\centerline{
\scalebox{0.3}{
\includegraphics{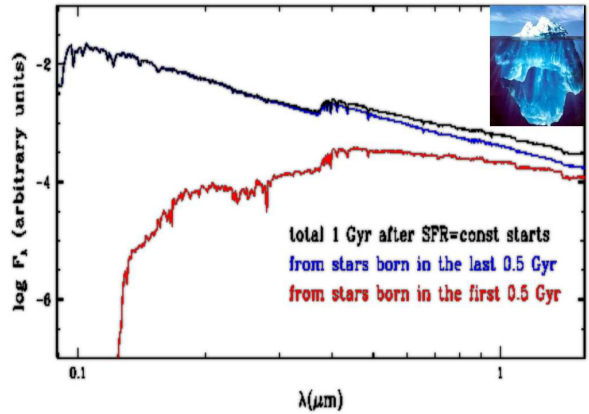}}}
\caption{The overshine effect of the youngest stars on the total galaxy spectrum. The total spectrum (black line) of a galaxy born 1 Gyr before and having formed stars constantly ever since is dominated by stars born in the last 0.5 Gyr (blue line) at virtually every wavelength. Only approaching the rest-frame near-IR their contribution become comparable to the one from stars born earlier (red line). Hence, the light we use for obtaining galaxy properties may be just the tip of the iceberg of the whole galaxy population. From Maraston et al. 2010.}
\label{overshine}
\end{figure}
\begin{figure}
\centerline{
\scalebox{0.4}{
\includegraphics{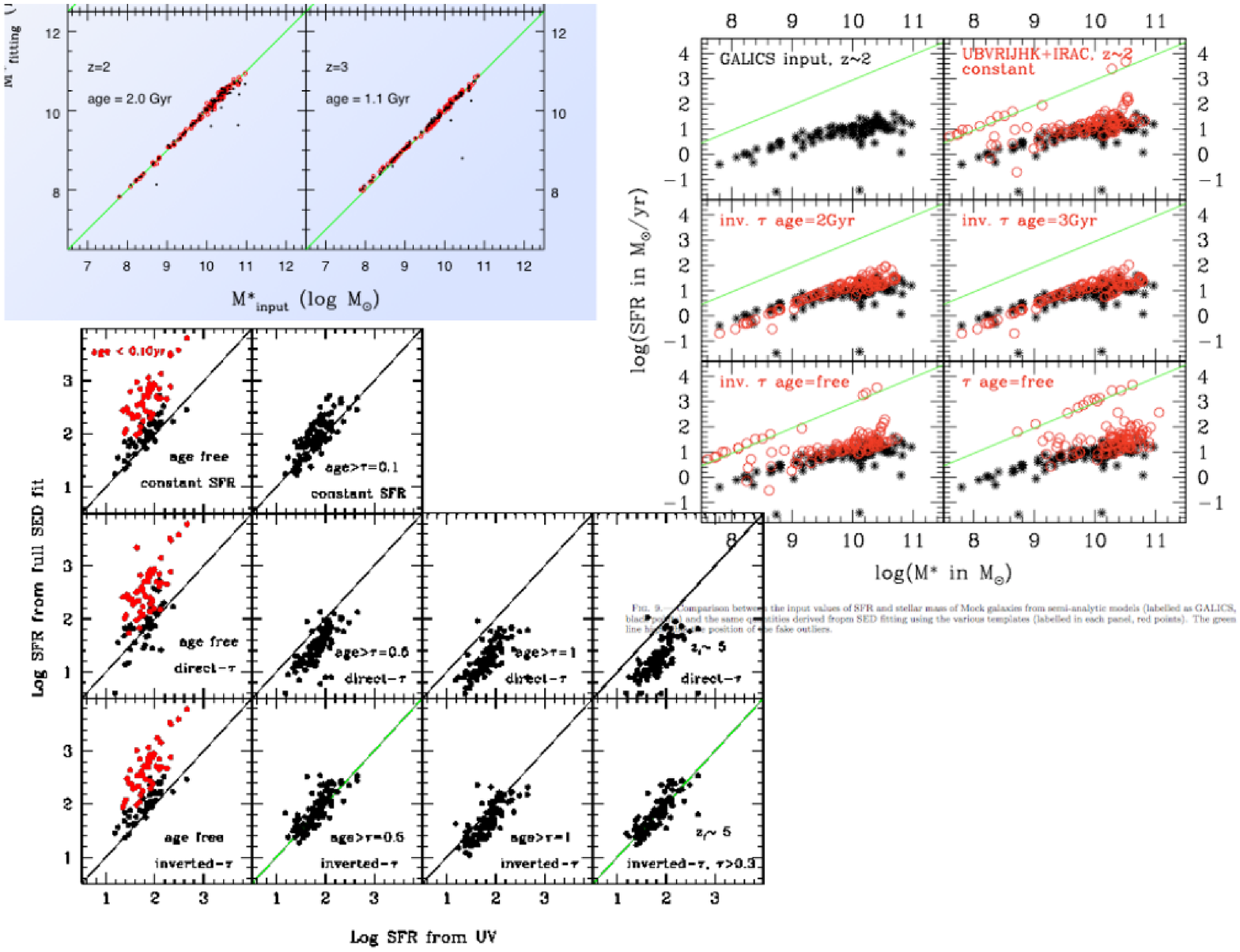}}}
\caption{Constraining the star formation mode of high-redshift galaxies. Left-hand plot: {\it calibration} of the SFR obtained fitting the full optical-to-near-IR (rest-frame) SED with models assuming different star-formation histories (namely exponentially-declining, constant or exponentially-increasing SF, which we call {\it inverted}$-\tau$ models; $y$-axis). The $x$-axis shows the SFR obtained with the $UV$-slope method, which is held to be an objective indicator of the true instantaneous SF (but see V. Buat's contribution for caveats). The winning model is shown in the lower-right panel. The right-hand plot shows a similar calibration, using now Mock galaxies with a known SFR vs mass relation (black points). Red points are the same galaxies after their photometry is interpreted with  various SFHs. The central row highlights how the best recovery of the input physical quantities is obtained with the exponentially-increasing models with high formation redshift. Both plots are from Maraston et al. 2010. Uppermost small plot: input mass vs mass from the fitting for the same Mock galaxies. The mass recovery using an appropriate template can be perfect (from J. Pforr et al., 2011, {\it submitted}).}
\label{inverted}
\end{figure}

While we leave the whole discussion around the modelling of the TP-AGB out of this contribution, as it has been discussed extensively in many other publications (see e.g. Maraston et al. 2011 and references therein), we note that in this redshift range the cosmic star formation peaks and galaxies are often find in a star forming mode (e.g. Renzini 2009; F\"orster-Schreiber et al. 2010). 

As galaxy physical properties are obtained by modelling the light emission, the case of star forming galaxies is particularly complicated. Indeed, the light of a galaxy is always dominated by the youngest populations, for the simple reason that the most massive stars have luminosities that are orders of magnitude larger than those of their least massive siblings, but their mass contribution can be quite small. (see Figure~\ref{overshine}). This fact strongly complicates the interpretation of model fits in case of star forming galaxies, as the doubt remains that - instead of obtaining properties for the whole galaxy - we may just be getting the tip of the iceberg. 

A recent paper provides a step forward towards the understanding of the mode of star formation of high-redshift star-forming galaxies. Maraston et al. (2010) fit redshift 2 galaxy data with several templates assuming different star formation histories, and various priors regarding the redshift at which galaxies started forming their stars and the typical timescale for star formation (see Figure~\ref{inverted}). They find that only one template is able to recover the star formation rate as derived by indicators independent of the model fitting, namely an exponentially-increasing star formation which started at a much higher redshift than the epoch of observations (e.g. $z\sim 5$) and has exponentially grown since with an e-folding time $\tau\sim0.4$ Gyr (Figure~\ref{inverted}, left-hand plot, lower right panel). All other setups, and in particular exponentially declining star formation histories - which are often assumed in the literature - are found to overestimate the SFR and underestimate the stellar mass. For example, the first column of the left-hand plot shows the results of a modelling where the age parameter is left free in the fitting. Because of the overshine effect (Figure~\ref{overshine}) the fit with the minimum $\chi^2$~has often a quite low age, which causes to derive a high SFR. Interestingly, the smaller SFRs and higher masses  of the best model could alleviate an existing discrepancy between the cosmic SFR and mass density (see M. Dickinson's review). The right-hand plot in Figure~\ref{inverted} shows that the same model is also able to recover at best the theoretical SFR vs mass relation of Mock galaxies from a semi-analytic model (middle row), and to perfectly recover their input stellar masses (uppermost small plot, showing the input mass vs the fitted mass). 
The bottom line is that the population modelling is a powerful tool to disclose galaxy evolution.
\subsection{Redshift below 1: modelling passive galaxies}

At redshift below 1, when a larger fraction of the galaxy population is old and nearly passive, the galaxy light is strongly contributed by Red Giant Branch stars (cf. Figure~\ref{starevol}). Here the main uncertainties of the theoretical modelling regard the temperature and the spectra of the cool RGB stars. The temperature is known to differ among various stellar tracks especially at high-metallicity (Maraston 2005), a regime that is relevant to massive galaxies (e.g. Thomas et al. 2010). The theoretical spectra are uncertain due to the low temperatures, and empirical spectra as used in recent models  have been shown to help in a better modelling of the data (Maraston et al. 2009). 

But also when a galaxy is nearly passive, the assumed star formation history matters. For example, Maraston et al. (2009) find that - for the $z\sim0.4$~Luminous (massive) Red Galaxies in Sloan - a model with old ages including a small fraction (by mass) of ancient metal-poor stars (Figure~6, right-hand plot) matches the median locus of the galaxy colours better than an old model with traces of residual star-formation (Figure 6 left-hand plot, see also Cool et al. 2008). The first model gets support by the fact that virtually all massive early-type galaxies at redshift zero possess a metal-poor halo (see e.g. Mehlert et al. 2003).
\begin{figure}
\centerline{
\scalebox{0.3}{
\includegraphics{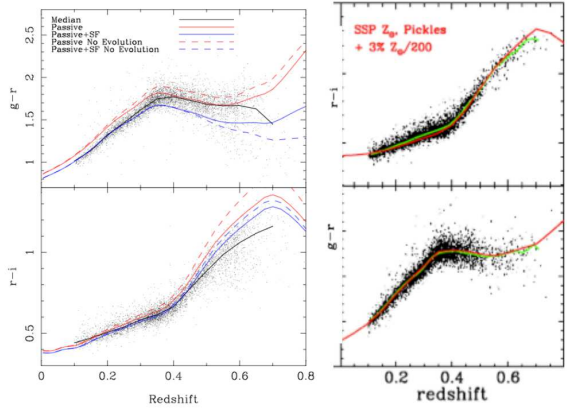}}}
\caption{Modelling the colour evolution of massive and nearly passive galaxies at $z<1$. A model with a small percentage of ancient metal-poor stars (right-hand plot, from Maraston et al. 2009) fit the data better than a model including residual star formation (blue line in the left-hand plot, from Wake et al. 2006), in agreement with the notion that local massive early-type galaxies possess a metal-poor halo.}
\label{lrg}
\end{figure}
There are evident astrophysical implications in choosing one or the other model, which we cannot discuss here, but more practically the stellar mass derived using the 'all old' model is higher than the one derived with a model including young stars, because the latter drives down the M/L (Tiret et al. 2011).

In conclusion, in this - quite wide - range of cosmic epochs the assumptions regarding the stellar evolution and the star formation history determine quite crucially the results obtained through data.
\section{Concluding Remarks: linking low to high-redshift}
\begin{figure}
\centerline{
\scalebox{0.4}{
\includegraphics{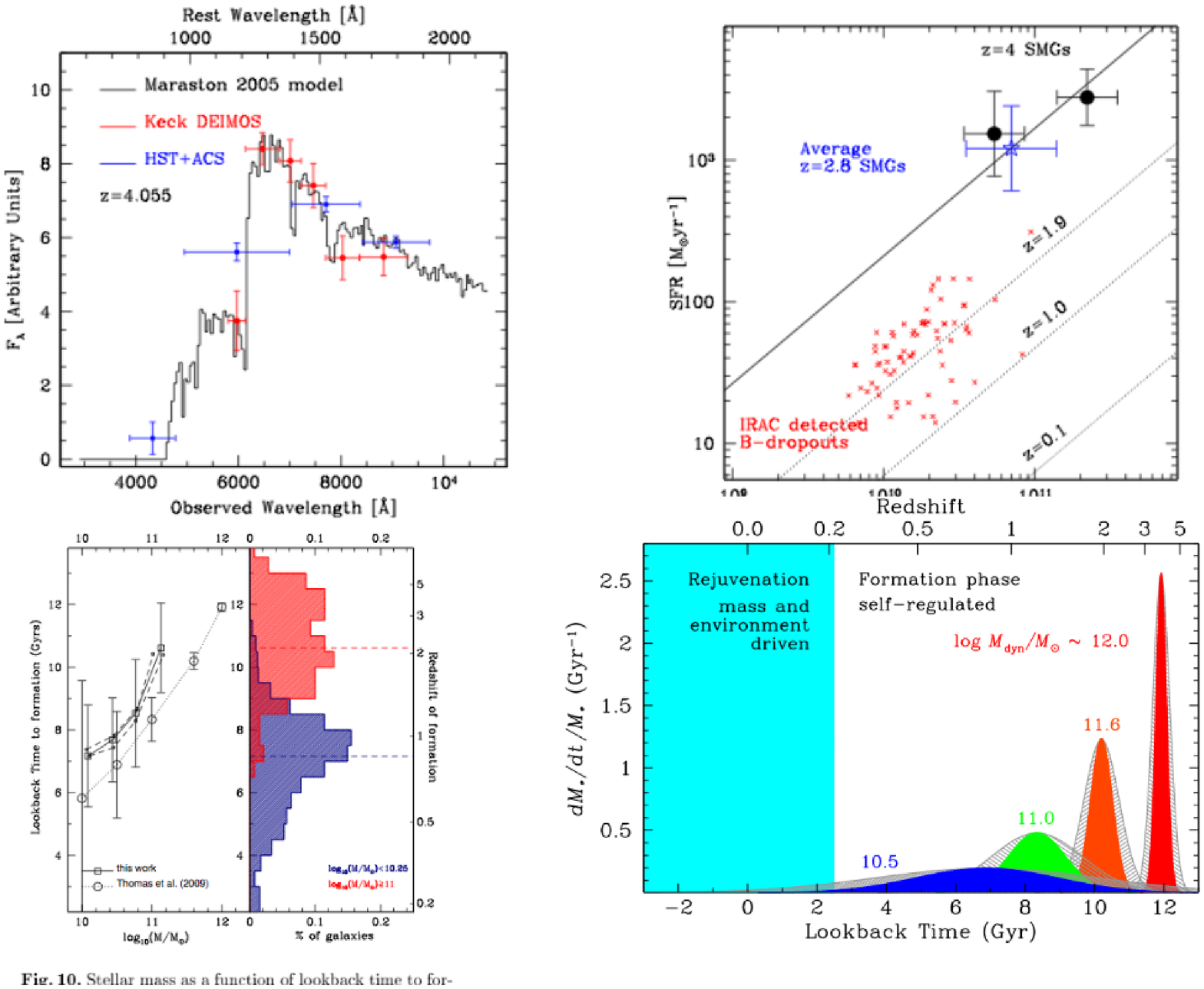}}}
\caption{Bridging low and high redshift galaxy evolution. Lower right-hand plot: The star formation histories of early-type galaxies as a function of their mass (Thomas et al. 2010), showing that the most massive galaxies formed at the highest redshifts and on the shortest timescales. Lower left-hand plot: a similar analysis performed at redshift 0.3 gives consistent results (Moresco et al. 2010). Upper plots, from Daddi et al. 2009: the analogue of local most massive galaxies could be the fast growing redshift 4 sub-millimeter galaxies with high SFRs.}
\label{link}
\end{figure}
The central goal of galaxy evolution studies is to understand the galaxy formation process in a cosmological context, and the primary challenge is to find the progenitors of nowadays galaxies. This is a wide research area, the results of which we cannot summarise in this paper (see Renzini 2006, Shapley 2011 for comprehensive reviews). Here I shall make some remarks on early-type galaxies, whose stellar populations and mass assembly have been traditionally challenging galaxy formation models (see reviews by C. Conselice, J. Silk, S. White, this volume). In particular, much is known about their stellar populations at redshift 0, which allows one to predict at which redshift their progenitors could be found. Figure~\ref{link} (right-hand plot) shows the star formation histories of local early-type galaxies as a function of their stellar mass, as derived from detailed chemical modelling of their spectra (Thomas et al. 2010). As well-known, the most massive galaxies formed at the highest redshift,  and a similar pattern is found at redshift $z\sim0.4$~(lower left plot, from Moresco et al. 2010; see also Le F\`evre, this volume). In particular $M^{*}\sim~10^{12}$~ should have formed around redshift 4. The upper plots show that interesting candidates as sub-millimeter galaxies (SMGs) at  redshift 4 have been found, whose high SFRs and epoch of detection support them as likely progenitors of nowadays massive galaxies (upper panels, Daddi et al. 2009). Instruments such as ALMA will be soon able to shed light on the early phases of massive galaxy formation (C. Casey, this volume). 

\end{document}